\documentclass[aps,amssymb,amsmath, pra, superscriptaddress]{revtex4}
\usepackage{graphicx}
\usepackage{epsfig}
\usepackage{dcolumn}
\usepackage{bm}
\usepackage{bbm}
\usepackage{hyperref}
\usepackage{color}
\usepackage[up]{subfigure}
\newcommand{\be}{\begin{equation}}
\newcommand{\ee}{\end{equation}}
\newcommand{\bc}{\begin{center}}
\newcommand{\ec}{\end{center}}
\newcommand{\bea}{\begin{eqnarray}}
\newcommand{\eea}{\end{eqnarray}}
\newcommand{\ba}{\begin{array}}
\newcommand{\ea}{\end{array}}
\begin{document}
\title{Relationship between quantum walks and relativistic quantum mechanics}
\author{C. M. \surname{Chandrashekar}}
\email{cmadaiah@iqc.ca}
\affiliation{Institute for Quantum Computing, University of Waterloo, 
Ontario N2L 3G1, Canada}
\affiliation{Perimeter Institute for Theoretical Physics, Waterloo, ON, N2L 2Y5, Canada}
\author{Subhashish Banerjee}
\email{subhashish@cmi.ac.in}
\affiliation{Chennai Mathematical Institute, Padur PO, Siruseri 603103, India}
\author{R. \surname{Srikanth}}
\email{srik@poornaprajna.org}
\affiliation{Poornaprajna Institute of Scientific Research,
Devanahalli, Bangalore 562 110, India}
\affiliation{Raman Research Institute, Sadashiva Nagar, Bangalore 560 080, India}


\begin{abstract}


Quantum walk models have been used as an algorithmic tool for quantum
computation and to describe  various physical processes.  This paper
  revisits the relationship between relativistic quantum mechanics and
  the  quantum walks.  We show  the similarities  of  the mathematical
  structure  of the decoupled  and coupled  form of  the discrete-time
  quantum  walk  to  that  of  the Klein-Gordon  and  Dirac  equations,
  respectively. In the  latter case, the coin emerges  as an analog of
  the  spinor  degree of  freedom.  Discrete-time  quantum  walk as  a
  coupled form  of the continuous-time  quantum walk is also  shown by
  transforming the decoupled form of the discrete-time quantum walk to
  the Schr\"odinger  form. By showing the  coin to be a  means to make
  the  walk  reversible,  and  that  the  Dirac-like  structure  is  a
  consequence of the coin use, our work suggests that the relativistic
  causal   structure    is   a   consequence    of   conservation   of
  information. However, decoherence (modelled by projective
  measurements on position space) generates entropy that increases with time, 
making the walk irreversible and thereby producing an arrow of time.
Lieb-Robinson bound is used to highlight the causal structure of the 
quantum walk to put in perspective the relativistic structure of quantum walk, 
the maximum speed the walk propagation and the earlier findings related to the 
finite spread of the walk probability distribution. We also present a two-dimensional 
quantum walk model on a two state system to which the study can be extended.
\end{abstract}

\maketitle
\preprint{Version}

\section{Introduction}

Quantum walks  are the  quantum analog of  the classical  random walks
\cite{Ria58,  FH65,  ADZ93} developed  using  the  aspects of  quantum
mechanics such as interference and superposition. Like their classical
counterpart, quantum  walks are also widely studied  in two forms:
continuous-time  quantum walk  \cite{FG98}  and discrete-time  quantum
walk \cite{ADZ93,  DM96, ABN01, NV01}. In the  continuous-time quantum walk,
one can directly  define the walk on the  {\it position} Hilbert space
$\mathcal{H}_p$,  whereas in  the  discrete-time quantum  walk, it  is
necessary to  introduce a quantum  coin operation, an  additional {\it
coin} Hilbert  space $\mathcal{H}_c$ to define the  direction in which
the particle  has to  evolve in position  space. The results  from the
continuous-time quantum  walk and  the discrete-time quantum  walk are
often  similar,   but  because of  the  coin  degree  of   freedom,  the
discrete-time  variant has  been shown  to be  more powerful  than the
other in some contexts \cite{AKR05}.
\par
Quantum walks  have emerged as a  powerful tool in  the development of
quantum  algorithms \cite{Amb03, CCD03,  SKB03}.  Furthermore,  it has
been  used to  demonstrate the  coherent quantum  control  over atoms,
quantum phase  transition \cite{CL08},  to explain phenomena  such as
breakdown of  an electric-field driven system  \cite{OKA05} and direct
experimental   evidence   for    wavelike   energy   transfer   within
photosynthetic    systems     \cite{ECR07,    MRL08}.     Experimental
implementation of the quantum walk has been reported with samples  
in nuclear magnetic resonance  system \cite{DLX03, RLB05};  in the continuous
tunneling of light fields through waveguide lattices \cite{PLP08};  in the phase space 
of trapped ions \cite{SMS09, ZKG10} based on the scheme proposed by \cite{TM02}; with single  optically trapped atoms \cite{KFC09}; and with single photon \cite{BFL10}.
Various other schemes  have been proposed for its physical realization in different physical systems \cite{RKB02, EMB05, Cha06, MBD06}.
\par
The relationship  between the idea  of quantum  walk and  relativistic quantum
mechanics  goes back to  the discrete  version of  the one-dimensional
Dirac   equation   propagator   considered   by  Feynman   and   Hibbs
\cite{FH65}.  Later, similarities of  the relativistic  wave equations
and unitary cellular and quantum  lattice gas automata were observed by
Bialynicki-Birula \cite{Bia94} and  Meyer \cite{DM96}.  Recently, that
is, after  the extensive  studies of the  one-dimensional discrete-time
quantum  walk,  the  relationship  between  quantum  walk  models  and
relativistic  quantum  mechanics  has  become a  topic  of
interest \cite{KFK05, Str06, BES07,  Str07, Kur08, SK10}.  In reference \cite{KFK05}, the one-dimensional quantum walk is mapped to the three-dimensional Weyl equation. In different continuum limits, the discrete-time quantum walk was shown to be equivalent to the one-dimensional Dirac equation and the continuous-time quantum walk, respectively \cite{Str06, Str07}. In reference \cite{BES07}, the evolved probability density for the Dirac particle was obtained from the asymptotic form of the probability distribution of the quantum walk. The effects similar to the relativistic effects, namely, Zitterbewegung and Klein's paradox,  were shown to be present in the discrete-time quantum walk \cite{Kur08}, and in reference \cite{SK10}, the Dirac equation with ultraviolet cutoff is shown to provide a discrete-time quantum walk in three dimensions on a four-component qubit \cite{SK10}.
\par 
The main focus  of this article is to present in  detail the dynamics of the
discrete-time  quantum   walk  and  the similarity   of  its  mathematical
structure to  that of  the relativistic quantum  mechanical evolution.
We compare the  similarities  of  the  mathematical  structures  of  the
decoupled and coupled  forms of the discrete-time quantum  walk to those 
of the  relativistic free spin-$0$ particles Klein-Gordon and free spin-$1/2$ 
particles Dirac equations, respectively. In  the latter
case,  the  coin  emerges  as  an  analog  of  the  spinor  degree  of
freedom.  By  showing  the  coin  to  be a  means  to  make  the  walk
reversible, and that the Dirac-like  structure is a consequence of the
coin use, our work suggests  that the relativistic causal structure is
a  consequence of  conservation of  information. We  also  discuss the
origin  of time asymmetry  in the  projective measurement  on position
space producing  the arrow of  time and making the  walk irreversible. The arrow of time in the quantum walk also been discussed recently \cite{SCS10}. Discrete-time quantum  walk as a  coupled form of  the continuous-time
quantum walk is  also shown by transforming the  decoupled form of the
discrete-time quantum  walk to the  Schr\"odinger form. The probability distribution for discrete-time
quantum  walk  evolution  spreads  in  time,  but  this  spreading  is
controlled  by  the coin  operation  used  during  the evolution.  The
presence of a speed limit in a discrete-time quantum walk dynamics is,
in  fact,  an instance  of  a much  more  general  phenomenon know  as the
Lieb-Robinson bound. Two-dimensional quantum walk using a two-state particle is presented, to 
which the study can be extended. 
\par 
In Sec. \ref{ctqwm},  we will  first review  the continuous-time
quantum  walk which follows  the Schr\"odinger  form of  evolution. In
Sec.  \ref{1dtqw}, we  will review  the discrete-time  quantum walk
model : In Sec. \ref{dsdtqw}, the dynamic structure of the discrete-time quantum walk is discussed, followed by the consequence of the projective measurement on the system, that is, the arrow of time, in Sec. \ref{arrowtime}. In Sec. \ref{rdtqw}, we will study the mathematical structure of the one-dimensional discrete-time quantum walk: its decoupled form and
similarities   to  the   free  spin-$0$   relativistic  form,   that  is,
the Klein-Gordon   form   (Sec. \ref{qwKG})  and   Schr\"odinger   form
(Sec. \ref{qwS}), and its coupled form and similarities to the Dirac form (Sec. \ref{dtqwdqf}).
In Sec.  \ref{LRqw}, we present  the Lieb-Robinson bound-like effect in
quantum  walk  evolution.   In  Sec.  \ref  {2dtqw}, a two-dimensional
discrete-time quantum walk model using a two-state particle, to which the study can be extended, is presented before concluding in Sec. \ref{conc}.

\section{Continuous-time quantum walk}
\label{ctqwm}
To  define the  continuous-time quantum  walk, it  is easier first to
define the  continuous-time classical random  walk and quantize  it by
introducing quantum amplitudes in place of classical probabilities.
\par 
The continuous-time classical random  walk takes place entirely in the 
position space.  To illustrate,  let us  define a continuous-time
classical random  walk on the position space  $\mathcal H_{p}$ spanned
by a vertex  set $V$ of a graph  $G$ with edge set $E$,  $G=(V, E)$. A
step of  the random walk  can be described  by an adjacency  matrix $A$
which  transforms the  probability  distribution over  $V$; that is,
 \bea
A_{j,k} = \begin{cases} 1 & ~~ (j,k) \in E, \\ 0 & ~~ (j,k) \notin E,
\end{cases}
\eea
for every pair $j, k \in V$.  The other important matrix associated with 
the graph $G$ is the generator matrix ${\bf H}$ given by
\bea
{\bf H}_{j,k} =  \begin{cases}
d_j \gamma   &   ~~ j =  k \\
-\gamma  &   ~~ (j,k) \in E \\
0  &  ~~ {\rm otherwise}
\end{cases}, 
\eea
where $d_{j}$  is the  degree of  the vertex $j$  and $\gamma$  is the
probability of transition between neighboring nodes per unit time.
\par
If $p_{j}(t)$ denotes  the probability of being at  vertex $j$ at time
$t$, then  the transition on  graph $G$ is  defined as the  solution of
the differential equation
\be
\label{ctcrw}
\frac{d}{dt} p_{j}(t) = - \sum_{k \in V}  {\bf H}_{j,k} p_{k}(t).
\ee
The solution of the differential equation is given by
\be
p(t) = e^{-{\bf H}t} p(0).
\ee
By replacing the probabilities $p_{j}$ by quantum amplitudes $a_{j}(t)
= \langle  j |  \psi(t) \rangle$ where  $|j\rangle$ is spanned  by the
orthogonal basis  of the position  Hilbert space $\mathcal  H_{p}$ and
introducing a factor of $i$ we obtain
\be
\label{ctqw}
i\frac{d}{dt} a_{j} (t) = \sum_{k \in V} {\bf H}_{j,k} a_{k}(t).
\ee
We can see that Eq. (\ref{ctqw}) is the Schr\"odinger equation
\be
i \frac{d}{dt} |\psi\rangle = {\bf H} |\psi \rangle.
\ee
Since the generator matrix is  an Hermitian operator, the normalization is
preserved  during  the dynamics.   The  solution  of the  differential
equation can be written in the form
\be
|\psi(t) \rangle = e^{-i{\bf H}t} |\psi(0)\rangle.
\ee
Therefore the  continuous-time  quantum  walk  is  of  the  form  of the 
Schr\"odinger equation, a nonrelativistic quantum evolution.
\par
To implement the continuous-time quantum  walk on a line, the position
Hilbert space  $\mathcal H_{p}$ can be  written as a  state spanned by
the basis  states $|\psi_{j} \rangle$,  where $j \in  \mathbb{Z}$. The
Hamiltonian ${\bf H}$ is defined such that
\be
\label{ctqw1b}     {\bf H}|\psi_{j}\rangle     =     
-\gamma |\psi_{j-1}\rangle     +
2 \gamma |\psi_{j}\rangle  - \gamma |\psi_{j+1}\rangle  
\ee 
and evolves the system through time $t$ via the transformation
\be
\label{ctqw1a}
U(t) =\exp(-i{\bf H}t).  \ee The  Hamiltonian ${\bf H}$ of the process
acts  as the  generator matrix  which will  transform  the probability
amplitude  at the  rate of  $\gamma$ to  the neighboring  sites, where
$\gamma$ is a time-independent constant.

\section{One-dimensional discrete-time quantum walk}
\label{1dtqw}

We   will  first   define   the  structure   of  the   one-dimensional
discrete-time  classical  random  walk.  The  discrete-time  classical
random walk takes place on the position Hilbert space $\mathcal H_{p}$
with instruction  from the  coin operation.  A  coin flip  defines the
direction in which the particle  moves, and a subsequent position shift
operation moves the particle in position  space. For a walk on a line,
a two-sided coin with a {\it  head} and a {\it tail} defines the movements
to the {\it left} and {\it right} side, respectively.
\par
The one-dimensional discrete-time quantum walk also has a very similar
structure  to that  of its  classical  counterpart. The  coin flip  is
replaced by the quantum coin operation which defines the superposition
of direction  in which the amplitude of the particle evolves simultaneously. The quantum
coin  operation followed by  the unitary  shift operation  is iterated
without  resorting to  intermediate measurement  to implement  a large
number of steps.  During the  walk on a line, interference between the
left-  and the  right- propagating  amplitude results  in  the quadratic
growth of variance with the number of steps.
\par
The discrete-time quantum walk on a line is defined on a Hilbert space
\be
\mathcal  H=  \mathcal H_{c}  \otimes \mathcal H_{p},
\ee
where $\mathcal H_{c}$  is the coin Hilbert  space and $\mathcal
H_{p}$  is  the position  Hilbert  space.  For a  discrete-time
quantum  walk in  one dimension,  $\mathcal H_{c}$  is spanned  by the
basis  state   (internal  state)  of  the   particle  $|0\rangle$  and
$|1\rangle$, and $\mathcal H_{p}$ is  spanned by the basis state of the
position $|\psi_{j}\rangle$,  where $j \in  \mathbb{Z}$.  To implement
the discrete-time quantum walk on a particle at origin in state 
\be
\label{qw:ins}
|\Psi_{in}\rangle= \left ( \cos(\delta)|0\rangle + e^{i\eta}
\sin(\delta)|1\rangle \right )\otimes |\psi_{0}\rangle,
\ee
the  quantum coin toss operation $B \in U(2)$, which in general can be 
written as  
\be 
\label{U2coin}
B_{\zeta, \alpha, \beta, \gamma} = e^{i \zeta} 
e^{i\alpha \sigma_{x}}e^{i\beta \sigma_{y}}e^{i\gamma \sigma_{z}},
\ee
is applied, where $\sigma_{x},  \sigma_{y}$, and $\sigma_{z}$  are the
Pauli  spin  operators.  Parameters  of the  coin  operations  $\zeta,
\alpha, \beta,  \gamma$ can be  varied to get  different superposition
states  of the  particle; that  is, quantum  coin  operation $B_{\zeta,
  \alpha,  \beta,   \gamma}$  is  used  to  evolve   the  particle  to
superposition  of  its basis  states  such that  it  can  serve as  an
instruction to  simultaneously evolve the  particle to the  $left$ and
$right$  of  its initial  position.   The  quantum  coin operation  is
followed by the conditional unitary shift operation $S$ given by
\bea
\label{eq:condshifta}
S &=& e^{-i(|0\rangle \langle 0|
- |1\rangle  \langle 1|)\otimes \hat{Pl}} = \left ( |0\rangle \langle
0|\otimes e^{-i\hat{P}l} \right )+ \left ( |1\rangle \langle 1|\otimes 
e^{i\hat{P}l} \right ),  
\eea
where $\hat{P}$ is  the momentum operator, $l$ is  the step length, and
$|0\rangle$   and   $|1\rangle$   are   the  basis   states   of   the
particle. Therefore the operator $S$, which delocalizes the wave packet
in  different  basis  states  $|0\rangle$  and  $|1\rangle$  over  the
position $(j-1)$ and $(j+1)$ when step length $l =1$, can also be
written as
\be
\label{eq:condshift}
S  =  |0\rangle  \langle 0|\otimes  \sum_{j  \in
\mathbb{Z}}|\psi_{j-1}\rangle  \langle \psi_{j} |+|1\rangle  \langle 1
|\otimes \sum_{j \in \mathbb{Z}} |\psi_{j+1}\rangle \langle \psi_{j}|.
\ee
The states in the new position are again evolved into the superposition
of  its basis state  and the  process of  quantum coin  toss operation
$B_{\zeta, \alpha, \beta, \gamma}$ followed by the conditional unitary
shift operation $S$,
\be
\label{dtqwev}
 W_{\zeta, \alpha, \beta, \gamma} =
S(B_{\zeta, \alpha, \beta, \gamma}  \otimes   {\mathbbm  1})
\ee
is iterated without resorting  to intermediate measurement, to realize
a large  number of steps of  the discrete-time quantum  walk. The four
variable parameters  of the quantum  coin, $\zeta, \alpha,  \beta$, and
$\gamma$ in Eq.  (\ref{U2coin}) can be  varied to change and  control the
probability amplitude distribution in the position space.
\par
The most widely studied form of the discrete-time quantum walk is the 
Hadamard walk, using the Hadamard operation $H = \frac{1}{\sqrt 2} 
\left( \begin{array}{clcr}
 1  & &   ~1   \\
1  & &  -1 
\end{array} \right)$ 
as a quantum coin operation, and the role of  the coin operation and initial
state  to  control the  probability  amplitude  distribution has  been
discussed  in   earlier  studies   \cite{ABN01,  BCG04}.  It   has  been
demonstrated that a three-parameter $SU(2)$ quantum coin operation,
\be
\label{3paraU2}
B_{\xi,\theta,\zeta} \equiv \left( \begin{array}{clcr}  
\mbox{~}e^{i\xi}\cos(\theta)  & &   e^{i\zeta}\sin(\theta)   \\
-e^{-i\zeta} \sin(\theta)  & &  e^{-i\xi}\cos(\theta)
\end{array} \right)
\ee
is sufficient to describe the most general form of the discrete-time 
quantum walk \cite{CSL08}.

\subsection{Dynamic structure of discrete-time quantum walk}
\label{dsdtqw}

The standard symmetric discrete-time classical random walk leads to
\be
\label{crwDE}
p(j,t+1) = \frac{1}{2}\left [ p(j-1,t)  + p(j+1,t) \right ], 
\ee 
where  $p(j,t)$ denotes  the probability  of finding  the  particle at
position  $j$  at  discrete  time  $t$. This  equation
expresses  the  fact that  all  the probability  at  a  given site  is
transmitted out at  each time step, so that  the probability available
at it in the next time step is that received in equal measure from its
immediate  neighbors. Subtracting  $p(j, t)$  from both
sides  of Eq. (\ref{crwDE})  leads to  the difference  equation which
corresponds to differential equation
\be
\label{eq:cla-dif}
\frac{\partial}{\partial  t} p(j,t) =  \frac{\partial^2}{\partial^2 j}
p(j,t),
\ee 
which is the standard classical diffusion equation. The preceding equation
is irreversible because the coin is effectively thrown away after each
toss.   It  is also  nonrelativistic  in the  sense  that  it is  not
symmetric in time  and space and leads to  a dispersion relation that
is   nonrelativistic   \cite{Str07}.   On   the   contrary,  in   the
discrete-time quantum walk, the information of the state of the quantum
coin in  the previous step  is retained and  carried over to  the next
step. This makes the quantum walk reversible.
\par
To  illustrate  this,  we  consider the  wave function  describing  the
position of a  particle and analyze how it evolves  with time $t$. Let
$t$ be the  time required to implement $t$ steps  of the quantum walk. The
two-component vector of amplitudes  of the particle, being at position
$j$, at  time $t$,  with left-moving ($L$) and right-moving  ($R$)  components, is
given by
\begin{eqnarray}
\label{compa0}
\Psi(j, t) = \left (  \begin{array}{cl} \Psi_L(j,t)  \\
\Psi_R(j,t)  \end{array} \right ). 
\end{eqnarray}
\par
A single-variable parameter quantum coin operation of the form 
\be
\label{qwcoin1}
B_{0, \theta, -\frac{\pi}{2}} = \left( \begin{array}{ccc}
\mbox{~~~}\cos(\theta) &~& -i\sin(\theta) \\
-i\sin(\theta) &~& \mbox{~~~}\cos(\theta) 
\end{array}\right)
\ee
is used  to drive the  dynamics for $\Psi(j,t)$.  The  coin parameters
$(0,  \theta,  -\frac{\pi}{2})$  have been  used  here  to  achieve  a
symmetrical form  of the coin  operation on the particle.  The quantum
coin operation is followed by the conditional shift operator $S$; that is, 
$S(B_{0,  \theta, -\frac{\pi}{2}} \otimes {\mathbbm  1})$ in terms
of left-moving ($L$) and right-moving ($R$) components at a given position
$j$ and time $t +1$ is given by
\begin{eqnarray}
\label{eq:compa2}
\left ( \begin{array}{cl} \Psi_L(j,t+1)  \\
\Psi_R(j,t+1)  \end{array} \right ) =  
\left( \begin{array}{ccc}
\mbox{~~~}\cos(\theta)a &~& -i\sin(\theta)a^{\dagger} \\
-i\sin(\theta)a ^{\dagger} &~& \mbox{~~~}\cos(\theta) a
\end{array}\right) \left ( \begin{array}{cl}  \Psi_L(j,t)  \\
\Psi_R(j,t)  \end{array}\right ), 
\end{eqnarray}
where action of operator $a$ and $a^{\dagger}$ on $\Psi(j,t)$ is given by
\begin{subequations}
\begin{eqnarray}
\label{eq:compa1}
a \Psi(j, t) = \Psi(j+1, t),  \\ 
a^{\dagger}\Psi(j, t) = \Psi(j-1, t).
\end{eqnarray}
\end{subequations}
Therefore,
\begin{subequations}
\begin{eqnarray}
\label{eq:compa}
\Psi_L(j,t+1) &=& \cos(\theta)\Psi_L(j+1,t) - 
	i \sin(\theta)\Psi_R(j-1,t),   \\
\label{eq:compb}
\Psi_R(j,t+1) &=& \cos(\theta)\Psi_R(j-1,t) - 
	i \sin(\theta)\Psi_L(j+1,t).
\end{eqnarray}
\end{subequations}
We thus  find that the coin  degree of freedom is  carried over during
the dynamics of the discrete-time quantum walk, making it reversible.

\subsection{Projective measurement, irreversibility  and arrow of time} 
\label{arrowtime}
From Eqs. (\ref{eq:compa}) and (\ref{eq:compb}), we noted that the coin
degree of freedom is carried over during the dynamics of the discrete-time
quantum  walk, making  walk  reversible; that  is,  the information  is
stored during  the evolution so that it can  be used to  reverse the
dynamics. However,  upon projective measurement on  the position space,
the information of the coin is lost, making the walk irreversible.  The
projective  measurement   produces  the   arrow  of  time   since  its
description  is time  asymmetric. Therefore an increase in measurement 
entropy of the system can be seen as an arrow of time.    
\par
\begin{figure}
\includegraphics[width=12.0cm]{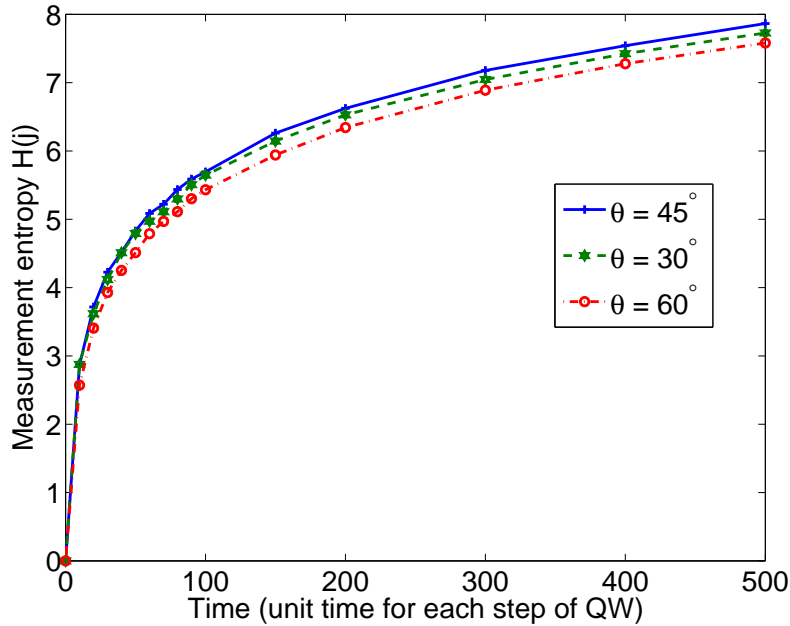}
\caption{\label{fig1}(Color online) Increase  in  measurement  entropy ($H(j)$)  with
  time when  unit time is required  to implement each  step of quantum
  walk. Increase  in entropy indicate  the arrow of  time. Measurement
  entropy was calculated for quantum walk  for up to 500 units in time
  using   $B_{0,   45^\circ,0}$,   $B_{0,  30^\circ,0}$   and   $B_{0,
    60^\circ,0}$ as quantum coin operations.}
\end{figure}

This  projective measurement of  position happening with  step time 1 can be understood  as an interaction with  the environment.  Quantum diffusion  via walk by  itself does  not generate entropy  (being unitary); rather interaction  with the  environment generates entropy  that  increases  with time. Figure (\ref{fig1})   depicts  the  increase   in  measurement  entropy with the increase in time. Measurement entropy is calculated by  considering the Shannon  entropy of the  particle position
probability  distribution  $p_j$ obtained  by  tracing  over the  coin
basis:
\be
H(j)= -\sum_{j=-t}^{+t} p_j \log p_j,
\ee
where $j$ is spanned over the position space at time $t$. 
This  time dependence  can  be understood  as  follows:  the position measurement generates entropy, which in each instance of measurement is translated into a classical record. That larger record of information is needed if the system is measured at a later rather than an earlier time, to reconstruct the original state. This can be construed as giving the direction of time. Following reference \cite{Mac09}, we may say that if the record for some process actually diminished along a direction of a time, there would be no objective knowledge of the process (here the walk) having happened.

\section{Relativistic features in discrete-time quantum walk} 
\label{rdtqw}

\subsection{Decoupled discrete-time quantum walk equation in Klein-Gordon form} 
\label{qwKG}

The discrete-time quantum walk can be written in a free spin-$0$ particles
relativistic form, that is, in the  Klein-Gordon form, by  decoupling the
components $\Psi_L$ and $\Psi_R$ in Eqs. (\ref{eq:compa}) and (\ref{eq:compb}) (Appendix \ref{AppendixA}):
\begin{subequations}
\begin{eqnarray}
\label{eq:deca}
\Psi_R(j,t+1) + \Psi_R(j,t-1) = \cos(\theta)[\Psi_R(j+1,t) + \Psi_R(j-1,t)], \\
\label{eq:decb}
\Psi_L(j,t+1) + \Psi_L(j,t-1) = \cos(\theta)[\Psi_L(j+1,t) + \Psi_L(j-1,t)].
\end{eqnarray}
\end{subequations}
Subtracting $2\Psi_R(j,t) + 2\cos(\theta) \Psi_R(j,t)$ from both sides
in  Eq.  (\ref{eq:deca}),  we   obtain  a  difference  equation  which
corresponds to differential equation
\begin{equation}
\label{eq:k-g}
\left [ \cos(\theta) \frac{\partial^2}{\partial j ^2} -  \frac{\partial^2}
{\partial t^2} \right ]\Psi_R(j,t) 
= 2(1-\cos(\theta))\Psi_R(j,t),
\end{equation}
by  setting  the  time step   and  spatial step  to  1;  see  Appendix \ref{AppendixB}  
for intermediate  steps. A similar
expression can be  obtained for $\Psi_L (j, t)$.  This shows that each
component follows a Klein-Gordon equation of the form
\begin{equation}
\left(\nabla^{2} - \frac{1}{c^{2}} \frac{\partial^{2}}{\partial t^{2}}\right ) 
\phi - \mu^{2}\phi = 0,
\end{equation}
showing essentially the free  spin-0 particles relativistic character of
each component of the discrete-time quantum walk.
\par
We obtain from  Eq. (\ref{eq:k-g}) the equivalent of  light speed $c$
and  mass $m$  of each  component $\Psi_{R}$  and $\Psi_{L}$  of the
discrete-time quantum walk dynamics :
\begin{eqnarray}
c &\equiv& \sqrt{\cos(\theta)},  \label{speed}\\
\mu= \frac{mc}{\hbar}&\equiv& \sqrt{2[\sec (\theta)-1]}.
\end{eqnarray}
Considering $\hbar =1$, we can write 
\begin{eqnarray}
 m &\equiv& \sqrt {\frac{2[\sec (\theta) -1]}{\cos (\theta)}}.
\end{eqnarray}
Note that  the maximum  velocity is given  by $c=1$,  corresponding to
$\theta=0$ and $m=0$, which  is in  agreement with  the relativistic
requirement that the rest mass  of light vanishes. This is also in agreement with the quantum walk dynamics that $\theta=0$ corresponds to state $|0\rangle$ and $|1\rangle$ moving away from each other without any interference, resulting in maximum variance \cite{CSL08}. The relativistic nature
of the quantum walk thus  arises as a natural consequence of employing
the  coin.   Since,  as  noted   earlier,  the  coin  makes  the  walk
reversible,  we have  the interesting  scenario that  the relativistic
causal  structure is  fundamentally a  consequence of  conservation of
information. This  is in accordance  with some recent works  that have
studied  possible  information theoretical  bases  for the  mathematical
structure of quantum mechanics \cite{CZ09,CBH03,SBG01,Sri06}.

\subsection{Decoupled discrete-time quantum walk equation in Schr\"odinger form} 
\label{qwS}

The  Klein-Gordon equation  can  be transformed  into the  Schr\"odinger
formulation  \cite{Gre}.  Transforming  the  discrete-time  equation,
Eq.  (\ref{eq:k-g})  -  which is  of  the  second  order in  the  time
coordinate - into a system  of two coupled differential equations that
are of first order in time is achieved by the ansatz
\be
\Psi_{R} = \varphi_{R} + \chi_{R} ~~~~~~,~~~~~  i\hbar 
\frac{\partial \Psi_{R}}{\partial t} = 
\sqrt {2[1 - \cos (\theta)]}~(\varphi_{R} - \chi_{R}),
\ee
in  which  $\Psi_{R}$  and  its  time  derivative  $\partial  \Psi_{R}
/  \partial  t$  are  expressed  as  components  of  two  functions
$\varphi_{R}$ and $\chi_{R}$.
\par
Now we can show that the two coupled differential equations,
\begin{subequations}
\begin{eqnarray}
\label{dtqw_se}
i\hbar\frac{\partial \varphi_{R}}{\partial t} = 
-\frac{\hbar^2}{2 \sqrt {\frac{2[\sec (\theta) -1]}{\cos (\theta)}} }~
\Delta (\varphi_{R} + \chi_{R}) + \sqrt {2[1 - \cos (\theta)]}~\varphi_{R},  \\
\label{dtqw_se1}
i\hbar \frac{\partial \chi_{R}}{\partial t} = \frac{\hbar^2}{2 
\sqrt {\frac{2[\sec (\theta) -1]}{\cos (\theta)}}} ~
\Delta (\varphi_{R} + \chi_{R}) - \sqrt {2[1 - \cos (\theta)]}~\chi_{R}, 
\end{eqnarray}
\end{subequations}
are  equivalent   to  the  discrete-time  quantum   walk  equation  in
relativistic form [Eq. (\ref{eq:k-g})] (Appendix \ref{AppendixC}).
\par
The coupled Eqs. (\ref{dtqw_se})  and (\ref{dtqw_se1}) may be combined
to form one equation by introducing the column vector
\be
\label{vs}
\Psi_{R} = \left ( \begin{array}{c}
\varphi_{R}  \\
\chi_{R} 
\end{array}\right)
\ee
and making use of the four $ 2 \times 2$ matrices
\be
\sigma_{1} = \left ( \begin{array}{ccc}
0  && 1  \\
1 && 0 
\end{array}\right), ~~~~~
\sigma_{2} = \left ( \begin{array}{ccc}
0  && -i  \\
i && 0 
\end{array}\right), ~~~~~~~~~
\sigma_{3} = \left ( \begin{array}{ccc}
1  && 0  \\
0 && -1 \end{array}\right),~~~~\mathbbm 1 = \left ( \begin{array}{ccc}
1  && 0  \\
0 && 1
\end{array}\right),
\ee
which satisfy the algebraic relations
\be
\sigma_{k}^{2}= \mathbbm 1 ~~,~~~~  \sigma_{k}\sigma_{l} 
= \sigma_{l}\sigma_{k} = i \sigma_{m} ~~,~~~~~{k, l, m = 1, 2, 3 -
{\rm in~a~cycle}}.
\ee
Using   the preceding relations,  we   can    combine   the   coupled
Eqs. (\ref{dtqw_se})  and (\ref{dtqw_se1})  to  form a  Schr\"odinger
-type equation, namely,
\be
\left ( i\hbar \frac{\partial}{\partial t } - {\bf \hat{H}_{R}} \right )  
\Psi_{R} = 0,
\ee
where ${\bf \hat{H}_{R}}$ is given by 
\be
{\bf \hat{H}_{R}} = \left ( \sigma_{3} + i \sigma_{2} \right ) 
\frac{\hat{P}^2 \sqrt{\cos (\theta)}}{2 \sqrt {2[\sec (\theta) -1]}} 
+ \sigma_{3} \sqrt{2[1 - \cos(\theta)]}.
\ee
Here  $\hat{P}  =  i  \hbar   \nabla$. Similarly  we  can  obtain  a
Hamiltonian ${\bf \hat{H}_{L}}$ for $\Psi_{L}$. Hence we have found that
the each component of the discrete-time quantum walk which has a structure
similar  to the  Klein-Gordon equation  can  be written  in a  coupled
Schr\"odinger equation formulation. Therefore a discrete-time quantum
walk can be  described as a coupled form  of a continuous-time quantum
walk   driven   by   Hamiltonians   ${\bf   \hat{H}_{R}}$   and   ${\bf \hat{H}_{L}}$.

\subsection{Discrete-time quantum walk equation in Dirac Equation form}
\label{dtqwdqf}
\par

In  Sec.  \ref{qwKG},  we  showed  that  the  decoupled  form  of  the
discrete-time quantum  walk equations leads  to a Klein-Gordon  form of
the  relativistic equation. Evolving  the discrete-time  quantum walk
equations  without decoupling,  that is,  in  a coupled  form, leads  to a
structure similar to $1+1$-dimensional Dirac equation:
\begin{eqnarray}
\label{de}
\left ( i \hbar \frac{\partial}{\partial t}  - {\bf \hat{H}_D}
 \right ) \Psi = \left ( i \hbar \frac{\partial}{\partial t}  + i \hbar c \hat{\alpha} \cdot \frac{\partial}{\partial x} - \hat{\beta} mc^2 \right ) \Psi = 0
\end{eqnarray}
where $m$ is the rest mass, $c$ is speed of light, $i\hbar \frac{\partial}{\partial t} $ is the momentum operator, and $x$ and $t$ are the space and time coordinates. The matrices $\hat{\alpha}$ and $\hat{\beta}$ are Hermitian and satisfy 
\be
\hat{\alpha}^2 = \hat{\beta}^2 = {\mathbbm 1}  ~~~~~~~~;~~~~~~~~ \hat{\alpha}\hat{\beta} = -\hat{\beta}\hat{\alpha}.
\ee
\par
To illustrate  this, we write  the coupled discrete-time  quantum walk
evolution equations [Eqs. \ref{eq:compa}) and (\ref{eq:compb}] in matrix form,
\begin{eqnarray}
\label{eq:compb1}
\left ( \begin{array}{cl} \Psi_L(j,t+1)  \\
\Psi_R(j,t+1)  \end{array} \right )  = \left [ \cos(\theta) 
\left( \begin{array}{ccc}
1 &~& 0 \\
0 &~& 1
\end{array}\right)  
\left ( \begin{array}{cl}  \Psi_L(j+1,t) \\  
\Psi_R(j-1,t)  
\end{array}\right ) \right ] + \left [ \sin(\theta) 
\left( \begin{array}{ccc}
0 &~& -i \\
-i&~& 0
\end{array}\right)   \left ( \begin{array}{ccc} \Psi_L(j+1,t)\\  
 \Psi_R(j-1,t)
\end{array}\right )\right ] .
\end{eqnarray}
The action of the coin operation ($B_{0, \theta, -\frac{\pi}{2}}$) and
condition sift operator ($S$) on  the particle commute with each other
for   the  discrete-time   quantum   walk  model   considered  in   the
article. Therefore the preceding expression can also be written in the form
\begin{eqnarray}
\label{eq:compc}
\left (  \begin{array}{cl} \Psi_L(j,t+1)  \\
\Psi_R(j,t+1)  \end{array} \right )  = B_{0, \theta, -\frac{\pi}{2}} 
\left ( \begin{array}{ccc} \Psi_L(j+1,t)\\  
  \Psi_R(j-1,t)
\end{array}\right )= \left [
 \cos(\theta) {\mathbbm 1} + \sin(\theta) \sigma_{3}\sigma_{2}  \right ] 
 \left ( \begin{array}{ccc} \Psi_L(j+1,t)\\  
  \Psi_R(j-1,t)
\end{array}\right ).
\end{eqnarray}
Subtracting both sides of the preceding equation by 
$\left (  \begin{array}{cl} \Psi_L(j,t)  \\
\Psi_R(j,t)  \end{array} \right ) + \left [
 \cos(\theta) {\mathbbm 1} + \sin(\theta) \sigma_{3}\sigma_{2} \right ]
 \left ( \begin{array}{ccc}  \Psi_L(j,t)\\  
 \Psi_R(j,t)
\end{array}\right )$ we get
\begin{eqnarray}
\label{eq:compd1}
\left (  \begin{array}{cl} \Psi_L(j,t+1)  -  \Psi_L(j,t) \\
\Psi_R(j,t+1)  -\Psi_R(j,t) \end{array} \right )  = \left [
 \cos(\theta) {\mathbbm 1}  + \sin(\theta) \sigma_{3}\sigma_{2} \right ]
 \left ( \begin{array}{ccc}  \Psi_L(j+1,t) - \Psi_L(j,t)\\  
 \Psi_R(j-1,t) - \Psi_R(j,t)
\end{array}\right )  
- \left (  \begin{array}{cl} \Psi_L(j,t)  \\
\Psi_R(j,t)  \end{array} \right )   \nonumber \\
+  \left [
 \cos(\theta) {\mathbbm 1}  + \sin(\theta) \sigma_{3}\sigma_{2} \right ]
 \left ( \begin{array}{ccc}  \Psi_L(j,t)\\  
 \Psi_R(j,t) 
\end{array}\right ).
\end{eqnarray}
The  difference form in  the preceding expression can  be reduced  to the
differential equation form
\begin{eqnarray}
\label{eq:compe}
\frac{\partial}{\partial t}   \left (  \begin{array}{cl} \Psi_L(j,t)  \\
\Psi_R(j,t)  \end{array} \right )
= \left [ \left (
 \cos(\theta) {\mathbbm 1} + \sin(\theta) \sigma_{3}\sigma_{2} \right )
  \left (  \begin{array}{cl} 
 ~~~ \frac{\partial}{\partial j} \\
 - \frac{\partial}{\partial j}
\end{array}\right )  
   + \left ( \cos(\theta) {\mathbbm 1} + \sin(\theta) 
\sigma_{3}\sigma_{2} - {\mathbbm 1} \right ) \right ]
\left ( \begin{array}{ccc}  \Psi_L(j,t)\\  
\Psi_R(j,t) 
\end{array}\right ).
\end{eqnarray}
By reordering and multiplying the entire expression by $i\hbar$, we obtain
\begin{eqnarray}
\label{eq:compe1}
i\hbar\frac{\partial}{\partial t}   \left (  \begin{array}{cl} \Psi_L(j,t)  \\
\Psi_R(j,t)  \end{array} \right )
= i\hbar \left [ \left (
 \cos(\theta) \sigma_{3} -  \sin(\theta) \sigma_{2}  \right )
 \frac{\partial}{\partial j}  + \left ( \cos(\theta) {\mathbbm 1} + 
\sin(\theta) \sigma_{3}\sigma_{2} - {\mathbbm 1} \right ) \right ]
\left ( \begin{array}{ccc}  \Psi_L(j,t)\\  
\Psi_R(j,t) 
\end{array}\right ).
\end{eqnarray}
When $\theta = 0$, the expression takes the form
\begin{eqnarray}
\label{eq:compe2}
\left [ i\hbar \frac{\partial}{\partial t} - i\hbar \sigma_{3} \frac{\partial}{\partial j} \right ]  \Psi(j,t) = 0.
\end{eqnarray}
The preceding expression is analogous to the $1+1$ dimensional Dirac equation of a massless particle $m=0$ in Eq. (\ref{de}).  Note  that $\theta=0$, and $m=0$ correspond to maximum velocity given by $c=1$. 
This is in agreement with both the relativistic  requirement   that  rest   mass   of  light   vanishes and the quantum walk dynamics with state $|0\rangle$ and state $|1\rangle$ moving away from each other without interfering, resulting in maximum variance \cite{CSL08}.
\\
From the Klein-Gordon form of discrete-time quantum walk discussed in Sec. \ref{qwKG}, we obtained $c\equiv  \sqrt{\cos(\theta)}$ and $mc^2  \equiv  \sqrt{2(1-\cos{\theta})}$.  In this section we have show that at certain limits, the discrete-time quantum walk structure is analogous to the Dirac equation of the massless particle. We note that effects similar to the Zitterbewegung effect and Klein paradox in the quantum walk have been studied using a different approach in \cite{Kur08}.

\section{Lieb-Robinson bounds in quantum walks} 
\label{LRqw}
\par
The probability distribution  for discrete-time quantum walk evolution
spreads  in  time,  but  this  spreading is  controlled  by  the  coin
operation used during the evolution.  The presence of a speed limit in
a nonrelativistic  dynamics is,  in fact, an  instance of a  much more
general  phenomenon.  Limits  to the  speed of  information propagation
known  as  Lieb-Robinson bounds  imply  that nonrelativistic  quantum
dynamics  has,  at least  approximately,  the  same  kind of  locality
structure provided in a field theory by the finiteness of the speed of
light.   The original work  by Lieb  and Robinson  pertaining to the bound
on the group velocity in quantum spin dynamics generated by a short-range
Hamiltonian dates back  to 1972 \cite{LR72}.  Since the work of Hastings \cite{Has04}, there have been a
series of extensions and improvements \cite{EO06, BV06, HK06, NS06, NO06} which show
that nonrelativistic quantum mechanics, with evolution governed by local 
Hamiltonians,  gives rise to an effective light cone with exponentially decaying tails.
This implies an emergence  of causality  in a quantitative manner in that the amount
of information exchanged between two regions not connected by a light cone
is exponentially small.
\par
The Lieb-Robinson bound states that the operator norm of the commutator of any
operators $O_A$ and $O_B$ in regions $A$ and $B$ at different times are
\begin{equation}
||[O_B(t),  O_A(0)]|| \leq C N_{min} ||O_A|| ||O_B|| \exp\left(-\frac{d_{AB} -  v t}{\kappa}\right), \label{lr}
\end{equation}
where $d_{AB}$ is the distance between $A$ and $B$; in graph theoretic terms the number
of edges in the shortest path connecting $A$ and $B$,  $N_{min} = $ min $\{|A|, |B|\}$, is the
number of vertices in the smallest of regions $A$ and $B$, while $C$, $v$, and $\kappa$ are 
positive constants depending upon the details of  the governing Hamiltonian  \cite{EO06, BV06}. 
\par
As an application of the Lieb-Robinson bound, [Eq. (\ref{lr})] to discrete-time quantum walk, let us consider  a one-dimensional quantum walk and take the operators $O_A$ and $O_B$ to be the square of the particle position, at the initial point before implementing quantum walk and at the end of $t$ steps of quantum walk  with unit time required to implement each step, respectively.  Bounds on  correlations are found from bounds on the corresponding commutator,  making use
of the Lieb-Robinson bound by making a spectral decomposition of the 
commutator and extracting the correlation as its negative frequency component \cite{Has04}.
The operator norms on the right-hand side of Eq. (\ref{lr}),  taken as the trace norm, 
would be the variance in the position of the particle at the initial point before starting the walk and at the end of $t$ steps of the walk. 
This would involve the probability distribution		
\begin{equation}
p(j,t) =  \langle j| \rm{Tr}_c \left(|\Psi_B^t \rangle \langle \Psi_B^t|\right)|j \rangle, \label{pdist}
\end{equation}
where $|\Psi_B^t \rangle = W_{\xi, \theta, \zeta}^t |\Psi_A \rangle$ is the state of the particle in position space  after $t$ steps of the walk,  with  $|\Psi_A \rangle$ referring to the initial state of the particle in position space,  Eq. (\ref{qw:ins}),  $W_{\xi, \theta, \zeta}$
is as in Eq. (\ref{3paraU2}),  and the trace operation in Eq. (\ref{pdist}) is the tracing over the
coin degrees of freedom.  
\par
For quantum walk using an unbiased coin operation, that is, $B_{\xi,\theta,\zeta}$ with $\xi = \zeta = 0$ , the variance after $t$ steps of quantum walk is $[1 - \sin(\theta)] t^2$ \cite{CSL08}. In Eq. (\ref{lr}),
$d_{AB}$ would be of the order of $t$, while $v$ would be $\sqrt{\cos(\theta)}$ as in Eq. (\ref{speed}). 
The Lieb-Robinson bound then tells us that the correlation function of the square of the particle position, for a $t$ step walk, is bounded above by $t^2$ and dies out exponentially beyond a region of the order of $t$. This is in accordance with \cite{ABN01, NV01},  where it was shown that for a quantum walk using a coin operator $B_{0,\theta,0}$ the probability distribution after $t$ steps is spread over the interval $[-t\cos(\theta), t\cos(\theta)]$ and quickly shrinks outside this region. The arguments using the Lieb-Robinson bounds thus put in perspective the preceding findings and also lend support to the causal structure of the quantum walk evolution brought out earlier by bringing out the relativistic features inherent in the quantum walk evolution, especially the connection to the Klein-Gordon equation and the identification of the corresponding velocity of quantum walk propagation [Eq. (\ref{speed})].

\section{Two-dimensional discrete-time quantum walk}
\label{2dtqw}

The description of  the discrete-time quantum walk on a line can be extended to the 2-D  plane by considering a two-state particle.  Operations can be defined on a two-state particle such that the amplitudes evolve in the $x$ direction and $y$ direction alternatively and show the relativistic structure in their evolution.
\par
For a two-dimensional discrete-time quantum walk on a plane using a two-state particle, the coin Hilbert space $\mathcal H_{c}$  is spanned  by the basis state (internal state) of the  particle $|0\rangle$ and $|1\rangle$ and the position Hilbert space $\mathcal H_{p}$  is spanned by the basis state of the position $|\psi_{j,k}\rangle$, where $j, k  \in \mathbb{Z}$ represent the two dimensions labeled by $j$ and $k$ elements in position space.
\par
The initial state of the two-state particle can be written as 
\be
\label{qw:ins3}
|\Psi_{in}\rangle= \left [ \cos(\delta)|0\rangle + e^{i\eta}\sin(\delta)|1\rangle \right ]\otimes |\psi_{0}\rangle.
\ee
It will be in a symmetric superposition state when $\delta = \pi/4$ and $\eta = \pi/2$.
\par
To realize a two-dimensional quantum walk using a two-state particle, a shift operator has to be constructed such that it will evolve the amplitudes of the particle in both the $x$ and $y$ directions. 
Therefore we will define the two shift operators $S_{x}$ and $S_{y}$ by
\begin{eqnarray}
\label{eq:2dxyShift}
S_{x}  &=&  |0\rangle  \langle 0| \otimes  \sum_{{j,k} \in \mathbb{Z}} |\psi_{j-1, k}\rangle  \langle \psi_{j, k} | + |1\rangle \langle 1| \otimes \sum_{j, k \in \mathbb{Z}} |\psi_{j+1, k}\rangle \langle \psi_{j, k} |,   \\
S_{y} &=& |\uparrow \rangle \langle  \uparrow |  \otimes  \sum_{{j,k} \in \mathbb{Z}}|\psi_{j, k-1}\rangle  \langle \psi_{j, k} |+ | \downarrow \rangle  \langle \downarrow | \otimes \sum_{j, k \in \mathbb{Z}} |\psi_{j, k+1}\rangle \langle \psi_{j, k}|,
\end{eqnarray}
where the relation between $|0\rangle$, $|1\rangle$, $| \uparrow \rangle$, and $| \downarrow \rangle$ is given by
\begin{eqnarray}
 |0\rangle =\frac{| \uparrow \rangle + | \downarrow \rangle}{2} , {\rm ~~~~~~~~} |1\rangle = \frac{|\uparrow \rangle - | \downarrow \rangle}{ 2} ,\nonumber \\ 
 | \uparrow  \rangle =|0\rangle + |1\rangle , {\rm ~~~~~~~~} | \downarrow \rangle = |0\rangle - |1\rangle.
\end{eqnarray}
If the particle is initially in the symmetric superposition state, 
\be
|\Psi_{ins} \rangle = \frac{1}{\sqrt 2}  \left ( |0\rangle + i | 1\rangle \right ) \otimes |\psi_{0,0}\rangle
\ee
 then,
 \begin{eqnarray}
S_{x}| \Psi_{ins}\rangle &=& \frac{1}{\sqrt 2} \left ( |0\rangle \otimes | \psi_{-1,0}\rangle + i |1\rangle  \otimes | \psi_{+1,0}\rangle \right ) = \frac{1}{2\sqrt 2}\left [ \left ( | \uparrow  \rangle + | \downarrow \rangle \right ) \otimes | \psi_{-1,0}\rangle + i  \left ( | \uparrow  \rangle - | \downarrow \rangle\right ) \otimes | \psi_{+1,0}\rangle\right ], \nonumber \\
S_{y}S_{x}| \Psi_{ins}\rangle &=& \frac{1}{ 2\sqrt 2} \left [ | \uparrow  \rangle \otimes | \psi_{-1,+1}\rangle + | \downarrow \rangle \otimes | \psi_{-1, -1}\rangle + i | \uparrow \rangle \otimes | \psi_{+1,+1}\rangle - i | \downarrow \rangle \otimes | \psi_{+1, -1}\rangle \right ] \nonumber \\
&=& \frac{1}{ 2\sqrt 2} \left [ \left ( | 0 \rangle + | 1  \rangle \right ) \otimes | \psi_{-1,+1}\rangle + \left (  | 0 \rangle - | 1 \rangle \right ) \otimes | \psi_{-1, -1}\rangle + i \left (  | 0 \rangle + | 1 \rangle \right )  \otimes | \psi_{+1,+1}\rangle \right. \nonumber\\  & & \left. - i  \left (  | 0 \rangle - | 1 \rangle \right )  \otimes | \psi_{+1, -1}\rangle \right ]. 
\end{eqnarray}
Therefore continuous iteration of $S_{y}S_{x}$ evolves amplitudes in superposition of position space implementing a two-dimensional quantum walk.
During operation $S_{x}$ the particle will evolve in the $x$ direction, and during operation $S_{y}$, the particle will evolve in  the $y$ direction; that is, if the order of operation is $S_{x}$ followed by $S_{y}$ then during every odd step, the evolution will be in the $x$ direction, and during every even step, the evolution will evolve the particle in the $y$ direction. 
The mathematical structure of the dynamics will be similar to that of the one-dimensional quantum walk and hence a relativistic structure similar to that of the one-dimensional quantum walk can be obtained. In higher dimensions,  the expression describing the evolution of the discrete-time quantum walk in $D$-dimensions can be decoupled to obtain a $2\times D$ number of expressions and worked out to be written in the relativistic forms. 

\section{Conclusion}
\label{conc}

In this article, we have shown the relationship between the mathematical structure of the discrete-time quantum walk and relativistic quantum mechanics. The dynamic structure of the one-dimensional discrete-time quantum walk using a two-sided coin quantum operation has been studied. The coupled structure of the dynamic expression of a discrete-time quantum walk is 
decoupled to arrive at an expression analogous to a free spin-0 particles, relativistic, Klein-Gordon form. We point out the quantum walk equivalents of the 
speed of light $c$ and mass $m$. We further use the same decoupled quantum walk expression 
to arrive at the Schr\"odinger formulation and show that the discrete-time quantum walk can be written as a coupled form of the 
continuous-time quantum walk. Furthermore, starting from a coupled form of the discrete-time quantum walk structure, we arrive at a 
mathematical structure analogous to the Dirac equation.  We have shown that the coin is a means to  make the  walk  reversible  and  that  
the  Dirac-like  structure  is  a consequence of the coin use.  Our work suggests that the relativistic  causal   structure    is   a   consequence    of   conservation   of   information. The existence of a maximum speed  of quantum walk propagation similar to the Lieb-Robinson  bound for  signal  propagation is also shown. 
This bound is used to highlight the causal structure of the walk and puts in perspective our work on the relativistic structure of quantum walk and earlier findings related to the finite spread of the walk probability distribution.

\vskip 1.00cm

\bc
{\bf Acknowledgement}
\ec
We thank S. Arunagiri, V. V. Sreedhar and R. Parthasarathy for discussions. CMC thanks Jens Eisert for stimulating conversation on Leib-Robinson bound. CMC also thank R. Simon and Institute of Mathematical Sciences, Chennai, India for hosting him during his multiple visits.

\appendix
\section{Decoupling the coupled expressions}
\label{AppendixA}

\underline{Getting Eqs. (\ref{eq:deca}) and (\ref{eq:decb}) from 
Eqs. (\ref{eq:compa}) and (\ref{eq:compb})}
\par
From Eq. (\ref{eq:compb}), solving for $\Psi_L$ we get
\begin{eqnarray}
\Psi_L(j+1,t) = \frac{i}{\sin(\theta)} \left [ \Psi_R(j, t+1)- \cos(\theta) 
\Psi_R(j-1, t) \right ],
\end{eqnarray}
\begin{eqnarray}
\Psi_L(j ,t+1) = \frac{i}{\sin(\theta)} \left [ \Psi_R(j-1,t+2) -  
\cos(\theta) \Psi_R(j-2,t+1) \right ].
\end{eqnarray}

Using this to substitution for $\Psi_L(j+1,t)$ and $\Psi_L(j ,t+1)$ in Eq. 
 (\ref{eq:compa}), we get Eq. (\ref{eq:deca}).
\par
From Eq. (\ref{eq:compa}), solving for $\Psi_R$, we get
\begin{eqnarray}
\Psi_R(j-1,t) = \frac{i}{\sin(\theta)}\left  [\Psi_L(j,t+1)- \cos(\theta) 
\Psi_L(j+1,t) \right ],
\end{eqnarray}
\begin{eqnarray}
\Psi_R(j ,t+1) = \frac{i}{\sin(\theta)} \left [ \Psi_L(j+1,t+2) - 
\cos(\theta) \Psi_L(j+2,t+1)\right ] .
\end{eqnarray}
Using this to substitute for $\Psi_R(j-1,t)$ and $\Psi_R(j ,t+1)$ in
Eq. (\ref{eq:compb}), we get Eq. (\ref{eq:decb}).

\section{Getting the difference operator that corresponds to the 
differential operators}
\label{AppendixB}

The difference operator $\nabla_t$ that corresponds to the differential
operator $\partial/\partial t$ is
\begin{eqnarray}
 \nabla_t = \frac{\Psi(j,t+\frac{h}{2}) - \Psi(j,t- \frac{h}{2})}{h}.
\end{eqnarray}
By setting the small incremental time to 1 ($h=1$), difference operator
\begin{eqnarray}
 \nabla_t = \Psi(j,t+ 0.5) - \Psi(j,t- 0.5)
\end{eqnarray}
corresponds to the differential operator $\partial/\partial t$.
Therefore the operator $\partial^2/\partial t^2$ will correspond to
applying the difference operator in each of the preceding two terms, which
yields
\begin{eqnarray}
 \nabla^2_t &=& \frac{1}{h} \times
\frac{[\Psi(j,t+1) - \Psi(j,t)] - [\Psi(j,t)-\Psi(j,t-1)]}{h}
\nonumber \\
&=& \frac{(\Psi(j,t+1) - 2\psi(j,t) +\Psi(j,t-1)}{h^2};
\end{eqnarray}
when  the  small  incremental  time  step  $h=1$,  it  corresponds  to
$\partial^2/\partial  t^2$.  The  difference operators  $\nabla_j$ and
$\nabla^2_j$    corresponding    to    $\partial/\partial    j$    and
$\partial^2/\partial j^2$ are also defined analogously for $j$, keeping
$t$ constant.

\section{Arriving at Klein-Gordon equation from two coupled 
equations}
\label{AppendixC}

This can be shown by subtracting Eq.  (\ref{dtqw_se1}) from Eq. (\ref{dtqw_se});

\begin{eqnarray}
\label{dtqw_se2}
i\hbar\frac{\partial }{\partial t} (\varphi_{R} - \chi_{R}) = 
-\frac{\hbar^2}{\sqrt {\frac{2[\sec (\theta) -1]}{\cos (\theta)}} }
~\Delta (\varphi_{R} + \chi_{R}) + \sqrt {2[1 - \cos (\theta)]}~(\varphi_{R} 
+ \chi_{R}), 
\end{eqnarray}

\begin{eqnarray}
\label{dtqw_se3}
i \hbar \frac{\partial }{\partial t} \left ( \frac{i\hbar} {\sqrt {2[1 - \cos (\theta)]}}\frac{\partial \Psi_{R}}{\partial t} \right ) = -\frac{\hbar^2}
{\sqrt {\frac{2[\sec (\theta) -1]}{\cos (\theta)}} }~\Delta \Psi_{R} + 
\sqrt {2[1 - \cos (\theta)]}~\Psi_{R},
\end{eqnarray}

\begin{eqnarray}
\label{dtqw_se4}
 \frac{1}{\sqrt {2[1 - \cos (\theta)]}} 
\frac{\partial^2 \Psi_{R}}{\partial t ^2} = \frac{1} 
{\sqrt {\frac{2[\sec (\theta) -1]}{\cos (\theta)}} }~
\Delta \Psi_{R} -  \sqrt {2[1 - \cos (\theta)]}~\Psi_{R} ,
\end{eqnarray}

\begin{eqnarray}
\label{dtqw_se5}
\frac{1}{\sqrt {2[\sec(\theta) - 1] \cos(\theta)}} 
\frac{\partial^2 \Psi_{R}}{\partial t ^2} = \frac{\sqrt{\cos (\theta)}}
{\sqrt {2[\sec (\theta) -1]}}~\Delta \Psi_{R} - \sqrt {2[\sec(\theta) - 1] 
\cos(\theta)}~\Psi_{R}. 
\end{eqnarray}

Thus we get:
\begin{eqnarray}
\label{dtqw_se6}
\left (\frac{1}{\cos(\theta)} \frac{\partial^2} {\partial t ^2} - \Delta 
\right ) ~\Psi_{R} = - 2[\sec(\theta) - 1]~\Psi_{R}. 
\end{eqnarray}

The preceding expression  is a recovery of the  discrete-time quantum walk
equation in Klein-Gordon form.


\end{document}